\documentclass[prl,twocolumn,showpacs,superscriptaddress]{revtex4}
\usepackage{graphicx}
\usepackage{amsmath}
\renewcommand{\section}[1]{{\par\it #1.---}}
\begin{document}
\title{Molecular Wires Acting as Coherent Quantum Ratchets}
\author{J\"org Lehmann}
\affiliation{Institut f\"ur Physik, Universit\"at Augsburg,
        Universit\"atsstr.~1, D-86135 Augsburg, Germany}
\author{Sigmund Kohler}
\affiliation{Institut f\"ur Physik, Universit\"at Augsburg,
        Universit\"atsstr.~1, D-86135 Augsburg, Germany}
\author{Peter H\"anggi}
\affiliation{Institut f\"ur Physik, Universit\"at Augsburg,
        Universit\"atsstr.~1, D-86135 Augsburg, Germany}
\author{Abraham Nitzan}
\affiliation{School of Chemistry, The Sackler Faculty of Science,
        Tel Aviv University, 69978 Tel Aviv, Israel}
\date{\today}
%
\begin{abstract}
The effect of  laser fields on the electron transport through
a molecular wire being weakly coupled to two leads is investigated.
The molecular wire acts as a coherent quantum ratchet if the molecule is composed of
periodically arranged, asymmetric chemical groups. This setup presents a quantum rectifier with a
finite dc-response in the {absence} of a static bias. The nonlinear current is evaluated
in closed form within the Floquet basis of the isolated, driven wire.
The current response reveals \textit{multiple} current
reversals together with a nonlinear dependence 
(reflecting avoided quasi-energy crossings) on both,
the amplitude and the frequency of the laser field.
The current saturates for long wires at a nonzero value, 
while it may change sign upon decreasing its length.

\end{abstract}
\maketitle

%
More than two decades after the idea of a molecular rectifier
\cite{Aviram1974a}, the experimental and theoretical study of such systems presently enjoys
a vivid activity \cite{Joachim2000a}.  Recent experimental progress has
enabled the reproducible measurement \cite{Cui2001a,Weber2001a} of weak
tunneling currents through molecules which are coupled by chemisorbed thiol
groups to the gold surface of the leads. A necessary ingredient for future
technological applications will be the possibility to control \textit{a
  priori} the tunneling current
through the molecule. Advances in that very direction have lately been achieved by
the development of a molecular field-effect transistor \cite{Schon2001a}.  
Typical energy scales in molecules are in the optical and the infrared regime,
where basically all of the today's lasers operate. Hence, lasers represent an
inherent possibility to control atoms or molecules and possibly currents
through them.  

A particularly intriguing phenomenon in strongly driven systems is the so-termed
ratchet effect \cite{Hanggi1996a,Astumian1997a,Julicher1997a,Reimann2002a},
originally discovered for overdamped classical Brownian motion in cyclic asymmetric
nonequilibrium systems. Counterintuitively to the second law one then observes 
a directed transport although all acting forces possess \textit{no} net bias. This effect has 
been established as well within the  regime of dissipative, incoherent quantum Brownian motion
\cite{Reimann1997a}. With this work we investigate the possibilities for 
quantum wires to act as a coherent
quantum ratchet, i.e.\ we consider the coherent quantum transport through  molecular wires with a
saw-tooth like level structure of the orbital energies when subjected to the influence of
a strong laser field.

Recent theoretical descriptions of molecular conductivity are based
on a scattering approach \cite{Mujica1994a,Datta1995a},
or assume that the underlying transport mechanism is an electron transfer
reaction from the donor to the acceptor site and that the conductivity can
be derived from the corresponding reaction rate \cite{Nitzan2001a}.
It has been demonstrated that both approaches yield identical results
 in a large parameter regime \cite{Nitzan2001b}.
Within the high-temperature limit, the electron transport on the wire
can be described by inelastic hopping events \cite{Nitzan2001a,Petrov2001a}.

Atoms and molecules in strong oscillating fields have been widely studied
within a Floquet formalism \cite{Manakov1986a,Grifoni1998a}. 
This suggests the following procedure: Making use of the tools that have
been acquired in that area, we
develop a formalism that combines Floquet theory for a driven molecule with
the many-particle description of transport through a system that is coupled to
ideal leads.  This approach is devised much in the spirit of the
Floquet-Markov theory \cite{Blumel1989a,Kohler1997a} developed for driven
dissipative quantum systems. Technically, we thereby work beyond the usual
rotating-wave approximation as frequently employed, such as e.g. in Ref.~\cite{Bruder1994a}.

\section{The lead-molecule model}
The total system composed of the driven wire, the leads, and the molecule-lead
couplings is described by the Hamiltonian
\begin{equation}
H(t)=H_{\text{wire}}(t) + H_{\text{leads}} + H_{\text{wire-leads}}
\end{equation}
The wire itself is modelled by $N$ atomic orbitals $|n\rangle$, $n=1,\ldots,N$,
which are in a tight-binding description coupled by hopping matrix elements.
The Hamiltonian for the electrons on the wire reads 
\begin{equation}
H_{\text{wire}}(t)=\sum_{n,n'} H_{nn'}(t)\, c_n^\dagger c_{n'},
\end{equation}
where the fermionic operators $c_n$, $c_n^\dagger$ annihilate,
respectively create, an electron in the atomic orbital $|n\rangle$.
The influence of the laser field is given by a periodic time-dependence
of the on-site energies yielding a single particle Hamiltonian of the
structure $H_{nn'}(t)=H_{nn'}(t+\mathcal{T})$, where
$\mathcal{T}=2\pi/\Omega$ is determined by the angular frequency $\Omega$ of the laser field.

\begin{figure}
\includegraphics[width=\columnwidth]{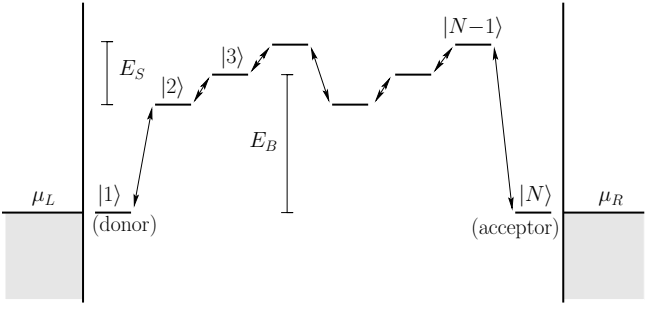}
\caption{\label{fig:wire}
Level structure of the wire ratchet with $N=8$ atomic sites, i.e.\
$N_g=2$ asymmetric molecular groups.
The bridge levels are $E_B$ above the donor and acceptor levels and are
shifted by $\pm E_S/2$.
}
\end{figure}

The orbitals at the left and the right end of the molecule, that we
shall refer to as donor and acceptor states, $|D\rangle=|1\rangle$ and
$|A\rangle=|N\rangle$, respectively, are coupled to ideal leads
(cf.\ Fig.~\ref{fig:wire}) by the tunneling Hamiltonian
\begin{equation}
H_{\text{wire-leads}}
=\sum_q ( V_{qL} \, c_{qL}^\dagger c_D + 
          V_{qR} \, c_{qR}^\dagger c_A) + \mathrm{h.c.},
\end{equation}
where $c_{qL}$ ($c_{qR}$) annihilates an electron in state $Lq$ ($Rq$)
on the left (right) lead.
In the following we will assume a wide-band limit
$\Gamma_{L/R} = (2\pi/\hbar) \sum_q |V_{qL/R}|^2 \delta(\epsilon-\epsilon_{qL/R})$,
i.e.\ an energy-independent coupling strength.

The leads themselves are modelled by grand-canonical ensembles of electrons with the
Hamiltonian $H_{\rm leads}=\sum_q(\epsilon_{qL}\, c_{qL}^\dagger c_{qL}+
\epsilon_{qR}\, c_{qR}^\dagger c_{qR})$
and electro-chemical potentials $\mu_{L/R}$.
Then the only non-trivial expectation values of lead operators are
$\langle c_{qL}^\dagger c_{qL}\rangle = f(\epsilon_{qL}-\mu_{L})$,
where $\epsilon_{qL}$ is the single particle energy of the state $qL$ 
and correspondingly for the right lead. Here, 
$f(\epsilon)=(1+e^{\epsilon/k_BT})^{-1}$ is the Fermi function.

\section{Perturbation theory}
While the dynamics of the leads and the wire, including the external driving, will be treated exactly,
we take the wire-lead Hamiltonian as a perturbation into account.
 From the Liouville-von Neumann equation $i \hbar \dot \varrho(t)=[H(t),\varrho(t)]$ for the total density
operator $\varrho(t)$ one obtains by standard techniques the master equation
\begin{align}
\label{mastereq}
\dot\varrho(t) = & -\frac{i}{\hbar}[H_{\rm wire}(t)+H_{\rm leads},\varrho(t)] \\
&-\frac{1}{\hbar^2}\int\limits_0^\infty \!d\tau
[H_\mathrm{wire-leads},[\widetilde H_\mathrm{wire-leads}(t-\tau,t),\varrho(t)]] . \nonumber
\end{align}
The tilde denotes operators in the interaction picture with respect
to the molecule and the lead Hamiltonian without the molecule-lead coupling,
$\widetilde X(t,t')=U_0^\dagger(t,t')\,X\,U_0(t,t')$, $U_0$ is the
propagator without the coupling.
The net (incoming minus outgoing) electrical current through the left contact
is given by the time-derivative of the electron number in the left lead
multiplied by the electron charge $-e$.  From Eq.~(\ref{mastereq}) follows in
the wide-band limit the expression
\begin{align}
I_L(t)  ={} & e \mathop{\rm tr}[\dot \varrho(t) N_L] = 
e\frac{\Gamma_L}{\pi\hbar}\mathop{\rm Re}\int\limits_0^\infty \!d\tau \! \int  \!d\epsilon\,
e^{i(\epsilon+\mu_L)\tau/\hbar}          
\nonumber\\
&\times
f(\epsilon)\big\langle[c_1,
\tilde c_1^\dagger(t-\tau,t)]_+\big\rangle 
-e\Gamma_L\big\langle c_1^\dagger c_1\big\rangle
\label{current_general}
\end{align}
and \textit{mutatis mutandis} for the net current through the right contact.
Equation (\ref{current_general}) expresses the current by the expectation
values $\langle[c_n,\tilde c_n(t-\tau,t)]_+\rangle$ and
$\langle c_n^\dagger c_n\rangle$.  We emphasize that these quantities depend on
the dynamics of the isolated wire and are thus influenced by the driving.

\section{Floquet decomposition}
Let us now focus on  the driven molecule
decoupled from the leads.
Since its Hamiltonian is periodic in time, $H_{nn'}(t)=H_{nn'}(t+\mathcal{T})$,
$\mathcal{T}=2\pi/\Omega$, we can solve the time-dependent Schr\"odinger equation
within a Floquet approach, i.e.\
we make use of the fact that there exists a complete set of solutions
of the form
$|\Psi_\alpha(t)\rangle=e^{-i\epsilon_\alpha t/\hbar}
|\Phi_\alpha(t)\rangle$ with the quasi-energies $\epsilon_\alpha$.
The so-called Floquet modes $|\Phi_\alpha(t)\rangle$ obey the
time-periodicity of the driving field and can thus be decomposed in a
Fourier series: $|\Phi_\alpha(t)\rangle=\sum_{k=-\infty}^\infty e^{-ik\Omega t}
|\Phi_{\alpha,k}\rangle$.
These obey  the quasienergy equation
\cite{Shirley1965a,Sambe1973a,Grifoni1998a}
\begin{equation}
\Big(\sum_{n,n'}|n\rangle H_{nn'}(t) \langle n'|-i\hbar\frac{d}{dt}\Big)
|\Phi_{\alpha}(t)\rangle = \epsilon_\alpha |\Phi_{\alpha}(t)\rangle .
\end{equation}

To make use of the knowledge about the driven molecule that
we obtain from Floquet theory, we define the Floquet representation
of the fermionic creation and annihilation operators by the time-dependent
transformation
\begin{align}
c_\alpha(t) &= \sum_n \langle\Phi_\alpha(t)|n\rangle\, c_n , \label{c_alpha}
\\
c_n &= \sum_{\alpha} \langle n|\Phi_\alpha(t)\rangle\,c_\alpha(t) .
\label{c_n}
\end{align}
The back transformation (\ref{c_n}) follows from the mutual orthogonality and
the completeness of the Floquet states at equal times \cite{Grifoni1998a}.  It
is now straightforward to prove that $\tilde
c_\alpha(t-\tau,t)=c_\alpha(t)\exp({i\epsilon_\alpha\tau/\hbar})$.  This
yields a spectral decomposition of Eq.\ (\ref{current_general}), which makes
it possible to evaluate the time and energy integrations therein.  Averaging
$I_L(t)$ over the driving period \cite{comment} leads to our first main
result, namely the average current
\begin{equation}
\begin{split}
\bar{I} = 
\frac{e\Gamma_L}{\hbar}\sum_{\alpha k}\Big[&
\langle \Phi_{\alpha, k}|D\rangle \langle D|\Phi_{\alpha, k}\rangle 
  f(\epsilon_\alpha+k\hbar\Omega-\mu_L) 
\\ &
-\sum_{\beta k'}
 \langle \Phi_{\alpha, k'+k} |D\rangle\langle D|\Phi_{\beta, k'} \rangle
 R_{\alpha\beta,k}
\Big] ,
\end{split}
\end{equation}
where $R_{\alpha\beta}(t)=\langle c_\alpha^\dagger(t) c_\beta(t)\rangle
=\sum_k e^{-ik\Omega t}R_{\alpha\beta,k}$.
Here we have used the fact that the $R_{\alpha\beta}(t)$ are expectation values
of a dissipative, periodically driven system.
Therefore, they share in the long-time limit the time-periodicity of the
driving field and can be represented by a Fourier series.

The remaining task in computing the stationary current is
to find the Fourier coefficients $R_{\alpha\beta,k}$ at asymptotic times.
For that purpose, we derive from the master equation
(\ref{mastereq}) an equation of motion for the $R_{\alpha\beta}(t)$.
Since all coefficients of the master equation as well as its
asymptotic solution are $\mathcal{T}$-periodic, we can split it into its
Fourier components which have to satisfy Eq.~(\ref{mastereq}) separately.
Finally, we obtain for the $R_{\alpha\beta,k}$ the inhomogeneous
set of equations
\begin{align}
\lefteqn{\frac{i}{\hbar}(\epsilon_\alpha-\epsilon_\beta+k \hbar \Omega)R_{\alpha\beta,k}}
\label{mastereq_fourier}
\\ &=&
 \frac{\Gamma_L}{2}\sum_{k'}
    \Big\{ &
    \sum_{\beta'k''}
                \langle\Phi_{\beta,k''+k'}|D\rangle
                \langle D|\Phi_{\beta',k''+k}\rangle
                R_{\alpha\beta',k'}
\nonumber \\&& {}+{} &
    \sum_{\alpha'k''}
                \langle\Phi_{\alpha',k''+k'}|D\rangle
                \langle D|\Phi_{\alpha,k''+k}\rangle
                R_{\alpha'\beta,k'}
\nonumber \\ && {}-{} &
    f(\epsilon_\alpha+k'\Omega-\mu_L)
    \langle\Phi_{\beta,k'-k}|D\rangle
    \langle D|\Phi_{\alpha,k'}\rangle
\nonumber \\ && {}-{} &
    f(\epsilon_\beta+k'\Omega-\mu_L)
    \langle\Phi_{\beta,k'}|D\rangle
    \langle D|\Phi_{\alpha,k'+k}\rangle
   \Big\}
\nonumber \\ &&&\hspace{-10ex} {} +
  {\big(\Gamma_L, \mu_L, |D\rangle\langle D|\big) \rightarrow
   \big(\Gamma_R, \mu_R, |A\rangle\langle A|\big)}.
\nonumber
\end{align}
For the typical parameter values used below,
a large number of side-bands contributes significantly to
the Fourier decomposition of the Floquet modes $|\Phi_{\alpha}(t)\rangle$.
Numerical convergence for the solution of the master equation
(\ref{mastereq_fourier}), however, is already obtained by using a few
side-bands for the decomposition of $R_{\alpha\beta}(t)$.
This keeps the numerical effort relatively small and justifies
the use of the Floquet representation (\ref{c_n}). Yet we are able
to treat the problem \textit{beyond} the usual rotating-wave-approximation
\cite{Bruder1994a}, which in certain parameter regimes turns out to
be crucial
\cite{tbp}.

\section{Ratchet wire}
We consider a molecular wire that consists of a donor and an acceptor
site and $N_g$ asymmetric molecular groups (cf.\ Fig.~\ref{fig:wire}).
Each of the $N=3N_g+2$ orbitals is coupled to its nearest neighbors by a
hopping matrix elements $\Delta$.
The laser field renders each level oscillating in time with a
position dependent amplitude.
Then the time-dependent wire Hamiltonian reads
\begin{equation}
H_{nn'}(t)=\left[E_n - A \cos(\Omega t)\, x_n\right]\delta_{nn'}
-\Delta(\delta_{n,n'+1}+\delta_{n+1,n'}) ,
\label{wirehamiltonian}
\end{equation}
where $x_n=(N+1-2n)/2$ is the scaled position of site $n$, the energy $A$ equals
electron charge multiplied by laser field strength and distance between two
neighboring sites, and $\Delta$ is the hopping matrix element.
The energies of the donor and the acceptor orbitals
are assumed to be at the level of the chemical potentials of the attached
leads and since no voltage is applied, $E_1=E_N=\mu_L=\mu_R$.  The bridge
levels $E_n$ lie at $E_B$ and $E_B\pm E_S/2$, respectively, as sketched in
Fig.~\ref{fig:wire}. We remark that for the sake of simplicity intra-atomic
dipole excitations are neglected within our model
Hamiltonian~(\ref{wirehamiltonian}). 

In our numerical studies, we use the hopping matrix element $\Delta$ as the
energy unit; in a realistic molecule, $\Delta$ is of the order $0.1\,{\rm
eV}$.  Thus, our chosen wire--lead hopping rate $\Gamma=0.1 \Delta/\hbar$
yields $e\Gamma=2.56\times10^{-5}$\,Amp\`ere and $\Omega=3\Delta/\hbar$
corresponds to a laser frequency in the infrared.  Note that for a typical
distance of {$2$\AA} between two neighboring sites, a driving
amplitude $A=\Delta$ is equivalent to an electrical field strength of
$2\times10^6\,\mathrm{V/cm}$.

\begin{figure}
\includegraphics[width=0.97\columnwidth]{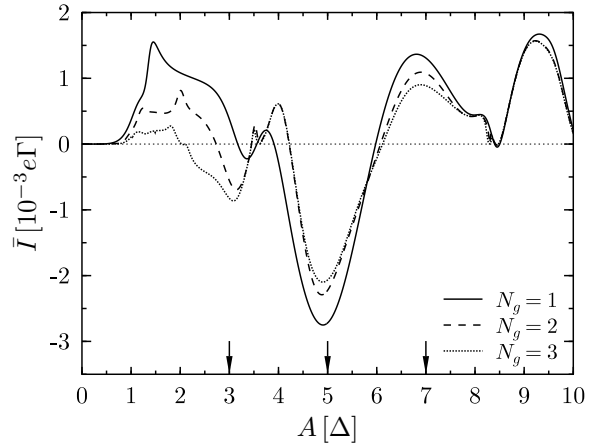}
\caption{\label{fig:I-F}
Time-averaged current (in arbitrary units) through a molecular wire that consists of $N_g$
bridge units as a function of the driving strength $A$.  The bridge 
parameters are $E_B=10 \Delta$, $E_S=\Delta$, the driving frequency is
$\Omega=3\Delta/\hbar$, the coupling to the leads is chosen as 
$\Gamma_L=\Gamma_R=0.1 \Delta/\hbar $, and the temperature
is $k_B T=0.25 \Delta$. 
The arrows indicate the driving amplitudes used in Fig.~\ref{fig:I-N}.
}
\end{figure}

\begin{figure}
\includegraphics[width=0.97\columnwidth]{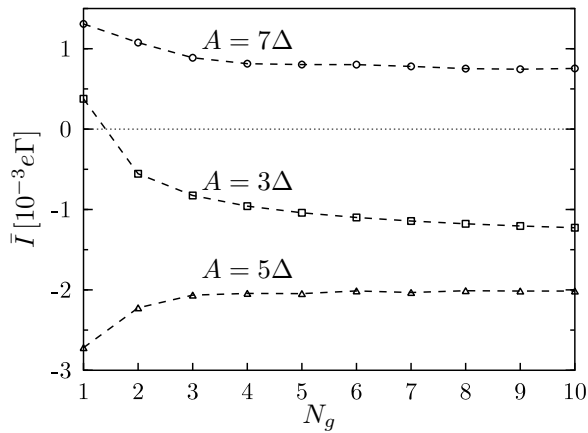}
\caption{\label{fig:I-N}
  Time-averaged current (in arbitrary units) as a function of the number of
  bridge units $N_g$ for the laser amplitudes indicated in Fig.~\ref{fig:I-F}.
  All other parameters are as in Fig.~\ref{fig:I-F}.  The connecting lines
  serve as a guide to the eye.  }
\end{figure}

Figure~\ref{fig:I-F} shows the stationary time-averaged current $\bar{I}$.
In the limit of a very weak laser field, we find $\bar{I}\propto E_S A^2$
(not shown).  This behavior is expected from symmetry considerations: The
asymptotic current must be independent of any initial phase of the driving
field and therefore is an even function of the field amplitude $A$.
This indicates that the ratchet effect can only be obtained from
a treatment beyond linear response.
For strong laser fields, we find that $\bar{I}$ is almost independent of the
wire length.  If the driving is intermediately strong, $\bar{I}$ depends
in a short wire sensitively on the driving amplitude $A$ and the number
of asymmetric molecular groups $N_g$: even the sign of the current may
change with $N_g$, i.e.\ we find a current reversal as a function of the
wire length.
For long wires that comprise five or more wire units, the average current
becomes again length-independent, as can be seen from Fig.~\ref{fig:I-N}. 
This identifies the observed current reversal as a finite size effect.
The fact that $\bar{I}$ converges to a finite value if the number of
wire units is enlarged, demonstrates that the dissipation caused by
the coupling to the leads is sufficient to establish the ratchet effect
in the limit of long wires.  In this sense, no on-wire dissipation
is required.
\begin{figure}
\includegraphics[width=0.97\columnwidth]{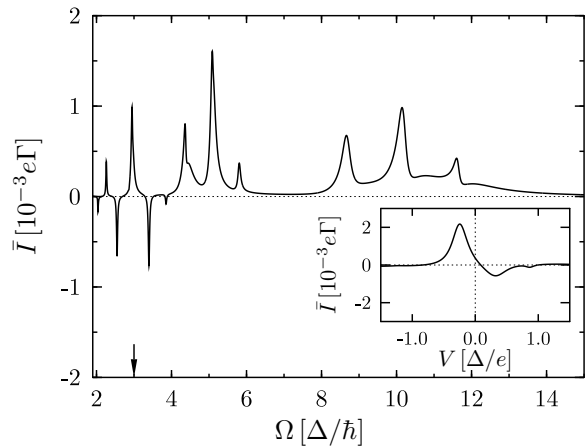}
\caption{\label{fig:I-Es}
  Time-averaged current (in arbitrary units) as a function of the driving
  frequency $\Omega$ for $A=\Delta$ and $N_g=1$.  All other parameters are
  as in Fig.~\ref{fig:I-F}. The inset displays the dependence of
  the average current on an externally applied static voltage $V$, which we
  assume here to
  drop solely along the molecule. The driving frequency is  $\Omega=3\Delta/\hbar$ 
  (cf. arrow in main panel).}
\end{figure}

Figure~\ref{fig:I-Es} depicts the average current \textit{vs.}\ the driving
frequency~$\Omega$, exhibiting resonance peaks as a striking feature.
Comparison with the quasi-energy spectrum reveals that each peak
corresponds to a non-linear resonance between the donor/acceptor
and a bridge orbital.
While the broader peaks at $\hbar\Omega\approx E_B=10 \Delta$ match the
1:1 resonance (i.e.\ the driving frequency equals the energy difference),
one can identify the sharp peaks for $\hbar\Omega\lesssim 7 \Delta$ as
multi-photon transitions. 
Owing to the broken spatial symmetry of the wire, one expects an asymmetric
current-voltage characteristic. This is indeed the case as depicted
in the inset of Fig.~\ref{fig:I-Es}. 

%
With this work we put forward an approach for the computation of the
current through a molecular wire in the presence of  laser fields of
arbitrary strength. Our method is based upon the Floquet solutions of
the isolated driven wire. The technique is  seemingly also
very efficient  for larger wire systems.
With this formalism at hand, we have established the possibility of
using molecular wires as coherent quantum ratchets.
As a finite size effect, the wire may exhibit a characteristic
current reversal as a function of (decreasing) length;
upon increasing the wire length the average current rapidly
converges to a finite value.
The strong dependence on the driving parameters in turn admits a tailored
quantum control of current with respect to both its sign and its magnitude. In
particular, this driven molecular wire with the distinctive multiple
current reversal feature encompasses new prospects to pump and shuttle
electrons on the nanoscale in an \textit{a priori} manner.  Our physical estimates show
that a realization of a molecular wire ratchet indeed is within experimental
reach.

This work has been supported by Sonderforschungsbereich 486 of the
Deutsche For\-schungs\-ge\-mein\-schaft
and by the Volkswagen-Stiftung under grant No.~I/77~217.

\bibliographystyle{prsty}


\end{document}